\documentclass[11pt]{article}
\usepackage[utf8]{inputenc}
\usepackage[T1]{fontenc}
\usepackage{lmodern}
\usepackage{amsmath, amssymb, amsfonts}
\usepackage{booktabs}
\usepackage{float}
\usepackage[margin=1in]{geometry}
\usepackage{hyperref}
\usepackage{natbib}
\usepackage{graphicx}
\usepackage{txfonts}
\usepackage{lineno}
\usepackage{algorithm}
\usepackage{algpseudocode} 
\usepackage{tikz}
\usetikzlibrary{
arrows.meta,
positioning,
shapes.geometric,
shapes.misc,
shapes.symbols,
calc,
fit,
decorations.pathreplacing,
backgrounds
}
\title{An Operator-Based Visual Analytics Pipeline for Synthetic Systemic Risk Dynamics}
\author{
Ana Isabel Castillo Pereda\\
\small Institute of Mathematics and Statistics, University of São Paulo, São Paulo, SP, Brazil\\
\small \textit{Correspondence:} anacp20@gmail.com; anaicp@ime.usp.br\\
\small Tel.: +55-11-983967981
}
\date{August 2026}

\begin{document}

\maketitle
\begin{abstract}

This work presents an operator-based visual analytics pipeline for exploring synthetic systemic risk dynamics in financial networks. The framework is formulated as a composition of mathematical operators that sequentially transform synthetic financial observations into dynamic scientific visualizations.

The pipeline consists of six operators: latent risk mapping, probabilistic score generation, financial network construction, distance-based contagion dynamics, visual encoding, and perspective projection. Together, these operators provide a modular computational structure linking nonlinear risk surfaces, time-dependent probabilistic states, network topology, and shock propagation.

A reproducible implementation demonstrates the architecture through controlled synthetic experiments. Nonlinear latent risk representations are converted into probabilistic scores, embedded in a weighted financial network, and propagated via shortest-path contagion. The resulting states are then mapped into dynamic visual representations. The experiments illustrate how visualization can be treated as an explicit stage of the analytical process rather than a post-processing step.

The proposed formulation does not aim to introduce new predictive models or contagion mechanisms. Instead, it offers a transparent and modular pipeline that connects generative risk modeling, network dynamics, and scientific visualization within a unified operator-based architecture. This approach supports reproducibility and facilitates the exploration and communication of complex systemic-risk processes in synthetic settings.

\noindent\textbf{Keywords:} Visual Analytics; Operator Composition; Synthetic Systemic Risk; Financial Networks; Network Contagion; Scientific Visualization; Computational Finance.
\end{abstract}

\section{Introduction}
\label{sec:introduction}

Modern financial systems are characterized by complex interactions among institutions, assets, and market participants. These interactions generate high-dimensional and dynamically evolving information whose analysis benefits from methodologies that combine network structure, stochastic processes, and risk assessment \cite{newman2010,Battiston2016}.

The importance of network topology for financial stability is well established. In a financial network $G=(V,E)$, institutions or assets are represented by nodes $V$, while contractual exposures, statistical dependencies, or other relationships define the edge set $E$. Systemic events typically emerge not only from the vulnerability of individual entities but also from the structure of their interconnections and the mechanisms through which distress propagates.

Several influential models have provided mathematical foundations for the analysis of such mechanisms. Eisenberg and Noe formalized equilibrium clearing payments among interconnected institutions \cite{Eisenberg2001}. Gai and Kapadia introduced threshold-based contagion in random financial networks \cite{gai2010}, while Lee incorporated liquidity shortages as an additional amplification channel \cite{lee2013}. Related work has further shown how asset-price feedback and network topology can transform localized shocks into system-wide instability \cite{cifuentes2005,acemoglu2015}.

More recent developments have extended this network perspective toward richer representations of systemic interactions. Contemporary research continues to emphasize the role of network structure in the analysis of systemic risk \cite{pacelli2025}, while multilayer formulations have shown how interactions across interconnected financial layers can generate additional mechanisms of risk amplification \cite{pangfan2024}. At the same time, graph-based learning methods are increasingly being investigated
as computational tools for estimating systemic-risk quantities directly from financial network structures \cite{gonon2026}. These developments illustrate a broader transition from isolated risk indicators toward computational representations in which probabilistic information, network structure, and propagation mechanisms can be analyzed jointly.

Despite progress in quantitative modeling, the exploration and communication of results in this domain still rely predominantly on static plots, summary tables, and isolated network diagrams. These representations remain essential, yet they often provide only partial insight into processes that are inherently temporal and interdependent, such as evolving risk scores, network contagion, and shock propagation.

Visual Analytics offers a complementary perspective by coupling analytical computation with interactive and dynamic visual representations designed to support the exploration of complex information \cite{thomas2005,keim2008}. Dynamic scientific visualization is particularly useful when the object of study itself evolves in time, allowing model states, network configurations, and propagation mechanisms to be examined within a common temporal framework.

In practice, however, risk modeling, network analysis, contagion simulation, and visualization are frequently treated as separate stages. Numerical outputs are generated first, network structures are characterized next, contagion is simulated independently, and visualization is typically introduced only at the end. As a result, the relationships among these components can remain fragmented even when the individual methods are sophisticated.

To address this fragmentation, the present paper introduces an operator-based visual analytics pipeline for synthetic systemic risk dynamics. The pipeline is expressed as the composition of mathematical operators

\begin{equation}
\mathcal{F}
=
\Pi
\circ
\Psi
\circ
\mathcal{C}
\circ
\mathcal{G}
\circ
\mathcal{P}
\circ
\mathcal{D},
\label{eq:pipeline_intro}
\end{equation}

where $\mathcal{D}$ constructs a latent risk representation from synthetic observations, $\mathcal{P}$ maps this representation into probabilistic scores, $\mathcal{G}$ builds the financial network, $\mathcal{C}$ evolves contagion dynamics using graph distance, $\Psi$ encodes quantitative states as visual attributes, and $\Pi$ performs the final projection onto the visualization domain. This formulation makes the information flow explicit while preserving the modularity of each stage.

The main contributions of this work are:

\begin{enumerate}
    \item the formulation of a modular operator-based architecture that connects latent risk mapping, probabilistic scoring, network construction, contagion dynamics, and scientific visualization;
    
    \item an explicit mathematical representation of the complete pipeline as a composition of operators from synthetic financial information to dynamic visual representations;
    
    \item a reproducible computational implementation based on controlled synthetic experiments, combining nonlinear risk surfaces, shortest-path contagion, and time-dependent visual encoding;
    
    \item the treatment of scientific visualization as an integral analytical stage rather than a purely post-processing step.
\end{enumerate}

The proposed pipeline is not intended to replace established quantitative models of systemic risk. Instead, it provides a transparent computational layer through which generative risk models, network structures, and contagion mechanisms can be composed, examined, and communicated within a coherent visual analytics framework. Because the operators are defined through compatible input and output spaces, individual components can be replaced or extended without redesigning the entire pipeline.

This work builds upon previous studies on systemic risk and contagion in equity markets \cite{pereda2025,pereda2026,castillopereda2026jaes}. While those contributions focused primarily on cascade dynamics and network analysis, the present paper generalizes the research direction toward a unified operator-based visual analytics architecture.

The remainder of the paper is organized as follows.
Section~2 develops the mathematical and computational framework.
Section~3 describes the algorithmic methodology.
Section~4 presents the experimental configuration and results.
Section~5 discusses the main implications, reproducibility,
and scope of the proposed approach.
Finally, Section~6 concludes the paper and outlines directions
for future research.

\section{Mathematical and Computational Framework}
\label{sec:framework}

\subsection{Latent Financial Representation}

The first stage of the pipeline constructs a latent risk representation from synthetic multidimensional observations. Let

\begin{equation}
X_i = (X_{1,i}, X_{2,i}) \in \mathbb{R}^2,
\label{eq:feature_vector}
\end{equation}

denote the feature vector associated with observation $i$, where the components are independently drawn from a standard normal distribution:

\begin{equation}
X_{1,i},\ X_{2,i} \sim \mathcal{N}(0,1).
\label{eq:feature_distribution}
\end{equation}

A nonlinear latent risk surface is defined by

\begin{equation}
z_i = 1.25 \sin(1.4 X_{1,i}) + 0.95 \cos(1.2 X_{2,i}) + 0.55 X_{1,i} X_{2,i} - 0.25 X_{1,i}^2.
\label{eq:latent_surface}
\end{equation}

This transformation introduces nonlinear interactions between the input features and generates heterogeneous risk regions while remaining computationally tractable. The resulting latent space is the collection

\begin{equation}
\mathcal{L} = \{ z_i \}_{i=1}^{N}.
\label{eq:latent_space}
\end{equation}

Equation~\eqref{eq:latent_surface} provides the initial representation from which all subsequent stages of the pipeline are derived.

\subsection{Probabilistic Scoring}

The latent representation is transformed into probabilistic scores through the logistic (sigmoid) mapping

\begin{equation}
p_i = \sigma(z_i) = \frac{1}{1 + \exp(-z_i)},
\label{eq:probabilistic_score}
\end{equation}

where $\sigma(\cdot)$ denotes the sigmoid function. The quantity $p_i$ is interpreted as the probability that observation $i$ belongs to a high-risk state.

Binary states are subsequently generated according to

\begin{equation}
Y_i \sim \mathrm{Bernoulli}(p_i).
\label{eq:bernoulli_state}
\end{equation}

To incorporate temporal evolution, a time-dependent perturbation is added to the latent surface:

\begin{equation}
z_i(t) = z_i + \varepsilon_i(t),
\label{eq:temporal_latent}
\end{equation}

where

\begin{equation}
\varepsilon_i(t) \sim \mathcal{N}(0, \sigma_t^2)
\label{eq:temporal_noise}
\end{equation}

and the noise amplitude $\sigma_t$ decreases over time. The corresponding dynamic probabilistic score is then given by

\begin{equation}
p_i(t) = \sigma\bigl(z_i(t)\bigr) = \frac{1}{1 + \exp\bigl(-z_i(t)\bigr)}.
\label{eq:dynamic_score}
\end{equation}

Equations~\eqref{eq:probabilistic_score}--\eqref{eq:dynamic_score} define the probabilistic component of the pipeline and supply the time-dependent risk scores used in the subsequent network and visualization stages.

\subsection{Financial Network Construction}

The probabilistic scores are embedded into a graph representation

\begin{equation}
G = (V, E, W),
\label{eq:financial_graph}
\end{equation}

where

\begin{equation}
V = \{v_1, \ldots, v_M\}
\label{eq:vertex_set}
\end{equation}

denotes the set of nodes (financial entities),

\begin{equation}
E \subseteq V \times V
\label{eq:edge_set}
\end{equation}

is the set of edges representing pairwise relationships, and

\begin{equation}
W = (w_{ij})
\label{eq:weight_matrix}
\end{equation}

is the matrix of non-negative edge weights.  

Each node is assigned a sector label and a vulnerability score. Edges are generated according to a probabilistic rule that increases the likelihood of connection between nodes belonging to the same sector. The resulting weighted graph provides the structural substrate on which contagion dynamics operate and constitutes the computational environment in which systemic disturbances propagate.

\subsection{Dynamic Contagion Process}

Let $v_0 \in V$ denote the initial shock source, selected as the node with the highest vulnerability score. Define

\begin{equation}
d_i = d(v_i, v_0)
\label{eq:graph_distance}
\end{equation}

as the shortest-path distance between node $v_i$ and the shock source on the graph $G$.

The temporal evolution of contagion intensity at node $i$ is modeled by

\begin{equation}
C_i(t) = S(t) \exp(-\lambda d_i),
\label{eq:contagion_intensity}
\end{equation}

where $\lambda > 0$ is a positive decay parameter and

\begin{equation}
S(t) = \max\left(0,\ \min\left(\frac{t - t_0}{\Delta t},\ 1\right)\right)
\label{eq:shock_intensity}
\end{equation}

denotes the normalized shock intensity. The function $S(t)$ increases linearly from zero to one over a prescribed time interval $[t_0,\ t_0 + \Delta t]$ and remains equal to one thereafter.

This formulation produces a continuous, distance-attenuated propagation of distress across the network. Larger values of $\lambda$ concentrate the contagion near the source node, while smaller values permit broader diffusion. Equation~\eqref{eq:contagion_intensity} therefore links network topology directly to the temporal evolution of systemic risk.


\subsection{Visual Mapping}

A central component of the proposed framework is the explicit
formalization of scientific visualization as an operator within the
computational pipeline. Rather than treating graphical representation as a post-processing step, the framework defines a visual mapping operator

\[
\Psi_t:
\mathcal{S}(t)
\longrightarrow
\mathcal{V}(t),
\label{eq:visual_mapping}
\]

where $\mathcal{S}(t)$ denotes the systemic state of the network at time $t$, including its topology, node vulnerabilities, probabilistic scores, and contagion intensities, and $\mathcal{V}(t)$ denotes the corresponding set of visual attributes.

More explicitly, the operator may be written as

\[
\Psi_t:
\left(G,\mathbf{v},\mathbf{p}(t),\mathbf{C}(t)\right)
\longmapsto
\left(\mathbf{r}(t),\mathbf{s}(t),
\boldsymbol{\alpha}(t),\boldsymbol{\chi}(t)\right),
\label{eq:visual_attributes}
\]

where $G=(V,E,W)$ is the financial network,
$\mathbf{v}=(v_1,\ldots,v_M)$ denotes the vector of node
vulnerabilities, $\mathbf{p}(t)$ contains the time-dependent
probabilistic risk scores, and $\mathbf{C}(t)=(C_1(t),\ldots,C_M(t))$ contains the contagion intensities.

The visual variables $\mathbf{r}(t)$, $\mathbf{s}(t)$,
$\boldsymbol{\alpha}(t)$, and $\boldsymbol{\chi}(t)$ represent,
respectively, node positions, node sizes, opacity levels, and color
encodings. These quantities provide graphical channels through which
changes in vulnerability, probabilistic risk, network structure, and
contagion intensity can be represented dynamically.

The operator $\Psi_t$ therefore establishes an explicit correspondence between quantitative model states and their graphical representation.
Importantly, $\Psi_t$ does not modify the underlying contagion dynamics or probabilistic scores; it maps those quantities into visual variables that facilitate their temporal and structural interpretation. Visualization is thus incorporated as a formal transformation within the pipeline while remaining analytically distinct from the mechanisms that generate the underlying systemic states.

\subsection{Perspective Projection}

The final stage of the visual transformation maps three-dimensional
graphical coordinates onto the two-dimensional visualization domain.
For this purpose, the framework introduces the projection operator

\[
\Pi:
\mathbb{R}^{3}
\longrightarrow
\mathbb{R}^{2}.
\label{eq:projection_operator}
\]

Let: 

\begin{equation}
\mathbf{r}_i(t)
=
\bigl(x_i(t),y_i(t),z_i(t)\bigr)
\in\mathbb{R}^{3}   
\end{equation}

denote the three-dimensional visual position associated with node $i$
after application of the visual mapping operator $\Psi_t$. The projected coordinates are defined by

\begin{equation}
\Pi\bigl(\mathbf{r}_i(t)\bigr)
=
\left(
x_i(t)\,\gamma\bigl(z_i(t)\bigr),
y_i(t)\,\gamma\bigl(z_i(t)\bigr)
\right),
\label{eq:perspective_projection}
\end{equation}

where the depth-dependent scaling factor is

\begin{equation}
\gamma(z)
=
\frac{1}{1+\beta z},
\qquad
\beta>0,
\label{eq:perspective_scaling}
\end{equation}

subject to the condition

\begin{equation}
1+\beta z>0
\label{eq:projection_condition}
\end{equation}

throughout the visualization domain.

Equivalently, the projected coordinates satisfy

\begin{equation}
x_i'(t)
=
x_i(t)\,\gamma\bigl(z_i(t)\bigr),
\qquad
y_i'(t)
=
y_i(t)\,\gamma\bigl(z_i(t)\bigr).
\label{eq:projected_coordinates}
\end{equation}

The scaling factor $\gamma(z)$ introduces a depth-dependent rescaling of the visual coordinates. Consequently, graphical elements located at different depths are represented at different apparent scales in the two-dimensional visualization.

The operator $\Pi$ acts exclusively on the geometric representation generated by $\Psi_t$ and does not modify the underlying probabilistic, network, or contagion states. The final visual state may therefore be written as

\begin{equation}
\mathcal{Y}(t)
=
\Pi\bigl(\mathcal{V}(t)\bigr),
\label{eq:final_visual_state}
\end{equation}

which completes the transformation from quantitative systemic states to their two-dimensional visual representation.

\subsection{Rendering Pipeline}

The complete framework is obtained by sequentially composing the operators introduced above. Starting from the synthetic input space $\mathcal{X}$, the information flow can be represented as :

\[
\mathcal{X}
\xrightarrow{\mathcal{D}}
\mathcal{L}
\xrightarrow{\mathcal{P}}
\mathcal{R}
\xrightarrow{\mathcal{G}}
\mathfrak{G}
\xrightarrow{\mathcal{C}}
\mathcal{S}(t)
\xrightarrow{\Psi_t}
\mathcal{V}(t)
\xrightarrow{\Pi}
\mathcal{Y}(t),
\label{eq:operator_flow}
\]

where $\mathcal{L}$ denotes the latent representation space,
$\mathcal{R}$ the probabilistic risk-score space,
$\mathfrak{G}$ the space of weighted financial networks,
$\mathcal{S}(t)$ the time-dependent systemic state,
$\mathcal{V}(t)$ the visual-attribute space, and
$\mathcal{Y}(t)$ the final two-dimensional visualization domain.

Accordingly, the complete transformation may be written as:

\begin{equation}
\mathcal{F}_t
=
\Pi
\circ
\Psi_t
\circ
\mathcal{C}_t
\circ
\mathcal{G}
\circ
\mathcal{P}_t
\circ
\mathcal{D},
\label{eq:complete_pipeline}
\end{equation}

so that

\begin{equation}
\mathcal{Y}(t)
=
\mathcal{F}_t(\mathbf{X}).
\label{eq:final_pipeline_output}
\end{equation}

This representation makes explicit the compatibility between consecutive stages of the pipeline: the output of each operator provides the quantitative or representational state required by the subsequent operator. The formulation is modular in the sense that an individual operator may be replaced by another mapping with compatible input and output spaces without changing the overall composition.

Algorithm~\ref{alg:pipeline_methodology} provides the computational realization of this operator sequence. Synthetic observations are first transformed into a nonlinear latent risk representation and probabilistic scores.
These quantities are associated with a weighted financial network, after which a shock is propagated according to shortest-path distance.
The resulting time-dependent states are encoded into visual attributes through $\Psi_t$ and subsequently mapped onto the two-dimensional visualization domain through $\Pi$.

The rendering stage therefore constitutes the terminal representation
layer of the computational pipeline. It does not alter the probabilistic or contagion mechanisms, but exposes their evolving states through coordinated graphical variables. This separation between quantitative dynamics and visual representation preserves the analytical structure of the underlying model while allowing the visualization layer to be modified independently.

The modular architecture also permits future extensions in which the
synthetic probabilistic component, network-construction mechanism, or
contagion operator is replaced by empirically calibrated alternatives.
Possible extensions include real financial observations, alternative
network models, learned graph representations, and more detailed
systemic-risk propagation mechanisms.

\subsection{Unified Operator Representation}

The complete visual analytics pipeline can be interpreted as a composition of mathematical operators acting on a sequence of compatible information spaces. Starting from the synthetic input state $\mathcal{X}$, the successive transformations generate latent representations, probabilistic risk scores, network structures, contagion states, visual attributes, and finally a two-dimensional scientific representation.

For a given time $t$, the complete transformation can be written as:

\begin{equation}
\boxed{
\mathcal{Y}(t)
=
\Pi\!\left(
\Psi\!\left(
\mathcal{C}\!\left(
\mathcal{G}\!\left(
\mathcal{P}\!\left(
\mathcal{D}(\mathcal{X})
\right)
\right);\,t
\right)
\right)
\right)
}
\label{eq:FIE}
\end{equation}

where $\mathcal{X}$ denotes the synthetic financial input space;
$\mathcal{D}$ defines the latent risk representation;
$\mathcal{P}$ generates probabilistic risk scores;
$\mathcal{G}$ constructs the weighted financial network;
$\mathcal{C}$ determines the time-dependent contagion state;
$\Psi$ maps quantitative states into visual attributes; and
$\Pi$ projects the resulting geometric representation onto the final
visualization domain $\mathcal{Y}$.

Equivalently, the architecture admits the compact operator representation

\begin{equation}
\boxed{
\mathcal{F}
=
\Pi
\circ
\Psi
\circ
\mathcal{C}
\circ
\mathcal{G}
\circ
\mathcal{P}
\circ
\mathcal{D}
}
\label{eq:FIE_operator}
\end{equation}

with

\begin{equation}
\mathcal{F}:
\mathcal{X}
\longrightarrow
\mathcal{Y},
\qquad
\mathcal{Y}(t)=\mathcal{F}(\mathcal{X};t).
\label{eq:FIE_mapping}
\end{equation}

The composition emphasizes that the proposed architecture is defined not by any individual risk, network, or contagion model, but by the structured information flow among compatible computational stages. Each operator performs a distinct transformation while exposing an output that can be consumed by the subsequent stage.

This separation provides the modularity of the framework. In particular, alternative probabilistic scoring mechanisms, network-construction rules, contagion models, or visual encodings may be incorporated provided that their input and output representations remain compatible with adjacent operators. The operator formulation therefore specifies the architecture of the pipeline independently of the particular synthetic mechanisms used in the present implementation.

\section{Algorithmic Methodology}
\label{sec:algorithmic_methodology}

The proposed operator-based visual analytics pipeline is implemented as a
modular computational procedure connecting synthetic data generation,
probabilistic risk scoring, financial-network construction, contagion
dynamics, visual mapping, and scientific rendering.
Algorithm~\ref{alg:pipeline_methodology} summarizes the complete
computational workflow.

\begin{algorithm}[H]
\caption{Operator-Based Visual Analytics Pipeline}
\label{alg:pipeline_methodology}
\small
\begin{algorithmic}[1]

\Require Number of synthetic observations $N$, number of time steps $T$,
network size $M$, contagion decay parameter $\lambda$, cascade threshold
$\tau$, and random seed
\Ensure Time-dependent systemic-risk states, contagion indicators, and
dynamic scientific visualization

\State Generate synthetic features
       $X_i=(X_{1,i},X_{2,i})$, $i=1,\ldots,N$

\State Compute the nonlinear latent risk representation
       $z_i \gets f(X_{1,i},X_{2,i})$

\State Compute baseline probabilistic risk scores
       $p_i \gets \sigma(z_i)$

\State Generate synthetic binary market states
       $Y_i \sim \mathrm{Bernoulli}(p_i)$

\For{$t=1,\ldots,T$}

    \State Generate temporal perturbations $\varepsilon_i(t)$

    \State Update the latent state
           $z_i(t) \gets z_i+\varepsilon_i(t)$

    \State Compute time-dependent probabilistic risk scores
           $p_i(t) \gets \sigma(z_i(t))$

    \State Compute aggregate probabilistic risk
           $\bar{p}(t) \gets
           \frac{1}{N}\sum_{i=1}^{N}p_i(t)$

\EndFor

\State Construct the weighted financial network
       $G=(V,E,W)$ with $|V|=M$

\State Assign sector labels and vulnerability scores
       $\nu_i$ to nodes

\State Generate weighted edges according to the prescribed
       sector-dependent connectivity rule

\State Select the initial shock source
       $v_0 \gets \arg\max_{v_i\in V}\nu_i$

\State Compute shortest-path distances
       $d_i \gets d_G(v_i,v_0)$ for all reachable $v_i\in V$

\For{$t=1,\ldots,T$}

    \State Compute normalized shock intensity $S(t)$

    \For{each node $v_i\in V$}

        \If{$v_i$ is reachable from $v_0$}
            \State Compute contagion intensity
            $C_i(t) \gets S(t)\exp(-\lambda d_i)$
        \Else
            \State $C_i(t) \gets 0$
        \EndIf

    \EndFor

    \State Compute cascade size
    \[
    K(t) \gets
    \sum_{i=1}^{M}
    \mathbf{1}\!\left\{C_i(t)>\tau\right\}
    \]

    \State Compute maximum contagion
    \[
    C_{\max}(t) \gets
    \max_{1\leq i\leq M} C_i(t)
    \]

    \State Compute average contagion
    \[
    \bar{C}(t) \gets
    \frac{1}{M}\sum_{i=1}^{M} C_i(t)
    \]

    \State Map quantitative states into visual attributes using $\Psi$

    \State Apply the projection operator $\Pi$

    \State Update dynamic visual elements and analytical indicators

    \State Render frame $t$

\EndFor

\State Store intermediate and final computational states for reproducibility

\State Export the resulting scientific visualization

\end{algorithmic}
\end{algorithm}

\begin{figure}[htbp]
\centering
\includegraphics[width=\textwidth]{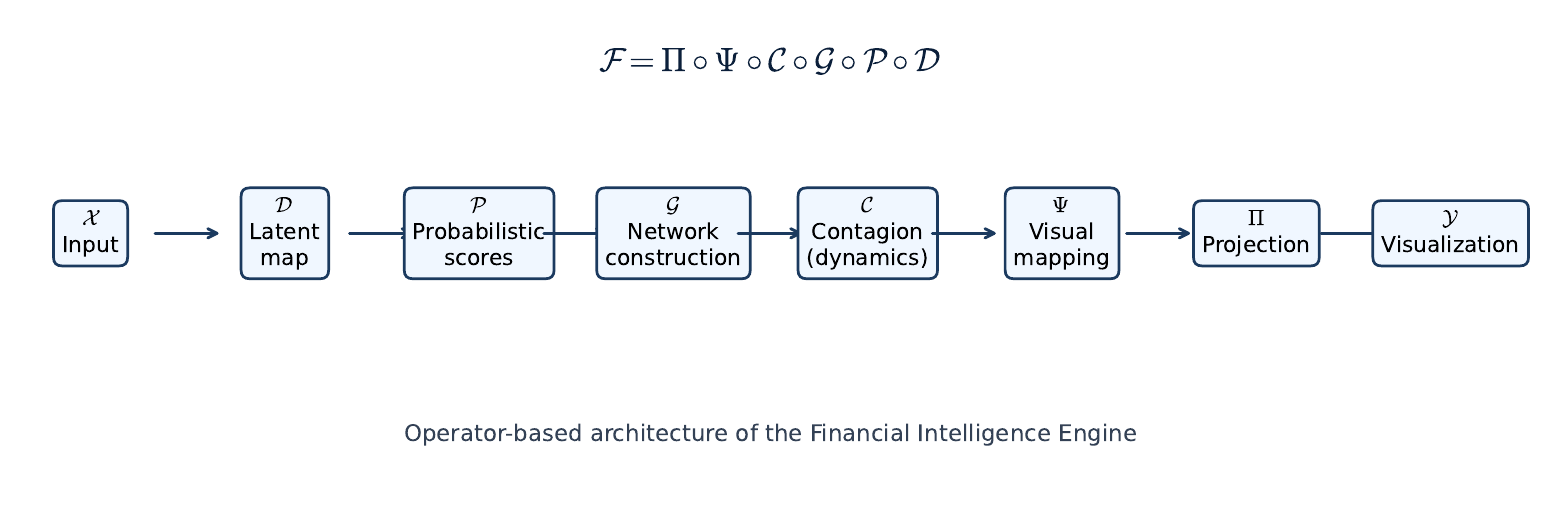}
\caption{
Operator-based architecture of the proposed visual analytics pipeline.
The framework is represented as the sequential composition of six operators:
latent risk representation ($\mathcal{D}$),
probabilistic risk scoring ($\mathcal{P}$),
weighted financial-network construction ($\mathcal{G}$),
shortest-path contagion dynamics ($\mathcal{C}$),
visual mapping ($\Psi$),
and projection onto the visualization domain ($\Pi$).
Starting from synthetic financial observations, the operators generate
progressively structured representations that connect probabilistic risk
states, network topology, contagion dynamics, and scientific visualization.
}
\label{fig:fie_architecture}
\end{figure}

Figure~\ref{fig:fie_architecture} summarizes the complete operator-based workflow proposed in this study.

Importantly, the latent risk representation is not constructed from
predefined discrete risk classes. Instead, its geometry is induced by the continuous nonlinear transformation defined in
Eq.~(\ref{eq:latent_surface}). The resulting latent values are
subsequently mapped into probabilistic risk scores through the logistic transformation,

\[
p_i = \sigma(z_i).
\]

Because both transformations are continuous, variations in the synthetic
input space produce gradual changes in the associated probabilistic risk
scores rather than an externally imposed hard partition of the observations.
Binary market states, when required by the synthetic experiment, are sampled
only subsequently from the corresponding Bernoulli distributions.

This construction establishes a direct transition from the latent
representation operator $\mathcal{D}$ to the probabilistic scoring operator
$\mathcal{P}$ while preserving the continuous structure of the synthetic
risk representation.

\begin{figure}[htbp]
\centering
\includegraphics[width=.9\textwidth]{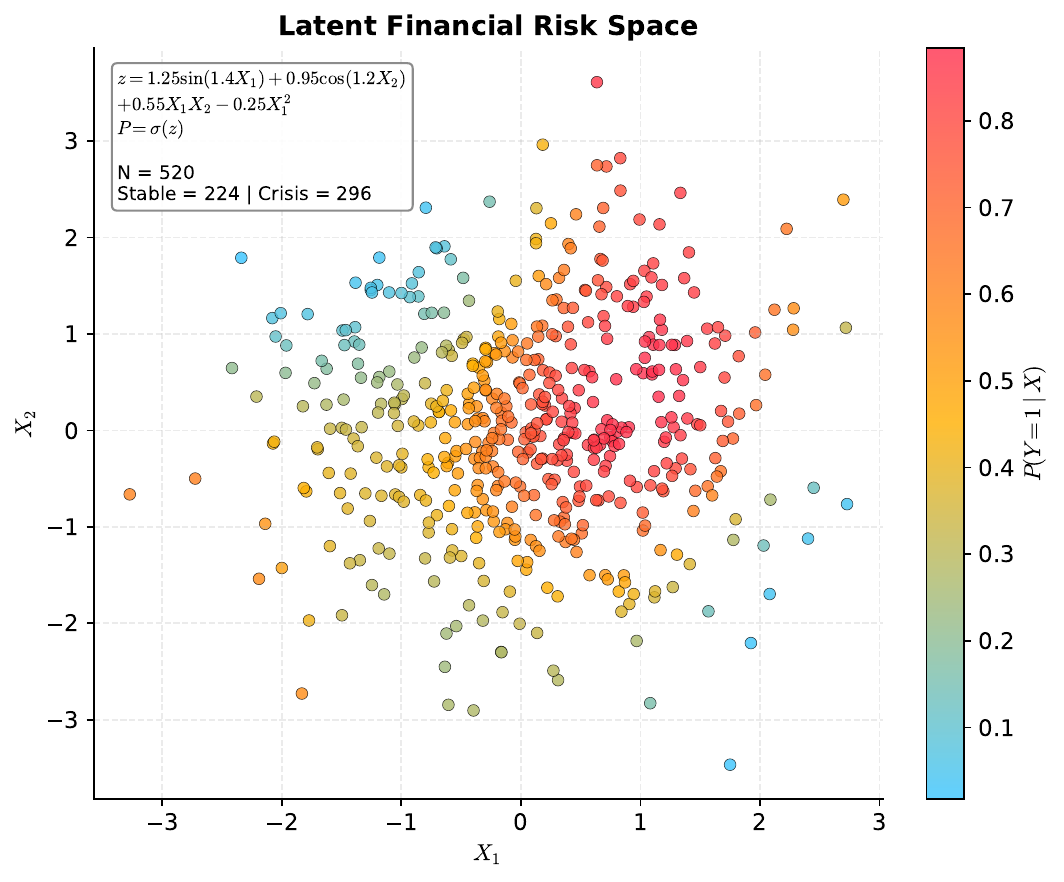}
\caption{
Synthetic latent risk representation induced by the nonlinear mapping
$\mathcal{D}$. Each point corresponds to one synthetic observation in the
two-dimensional feature space $(X_1,X_2)$. The vertical coordinate represents
the latent risk signal $z=f(X_1,X_2)$, while the Stable and Crisis markers
correspond to binary states sampled from
$Y_i\sim\mathrm{Bernoulli}(p_i)$, with $p_i=\sigma(z_i)$.
The embedded equation specifies the nonlinear transformation used in the
experiment.
}
\label{fig:latent_space}
\end{figure}

Figure~\ref{fig:latent_space} illustrates the nonlinear structure induced by
the latent representation operator $\mathcal{D}$. Synthetic observations
$(X_{1,i},X_{2,i})$ are mapped to latent values

\begin{equation}
z_i=f(X_{1,i},X_{2,i}),
\end{equation}

thereby producing a continuous risk surface over the two-dimensional input
space. The binary Stable and Crisis states shown in the figure are not used
to construct this surface. They are generated subsequently by sampling from
the Bernoulli distributions associated with the probabilistic scores
$p_i=\sigma(z_i)$.

The latent representation therefore remains continuous even though the
synthetic market states are binary. Regions associated with larger latent
values correspond to larger crisis probabilities after application of the
logistic transformation, whereas smaller latent values correspond to lower
probabilistic risk scores.

The probabilistic scoring operator $\mathcal{P}$ transforms the latent signal
according to

\begin{equation}
p_i=\sigma(z_i),
\end{equation}

mapping the nonlinear latent representation into the interval $(0,1)$.
Because the logistic function is strictly monotone, the ordering of latent
risk values is preserved under this transformation: if $z_i<z_j$, then
$p_i<p_j$.

From the operator-based perspective, Figure~\ref{fig:latent_space} therefore
illustrates the transition

\[
(X_{1,i},X_{2,i})
\xrightarrow{\mathcal{D}}
z_i
\xrightarrow{\mathcal{P}}
p_i,
\]

while the binary state $Y_i$ constitutes a subsequent stochastic realization
of the corresponding probability. Any probability contour used for visual
interpretation, such as $p=0.5$, should consequently be understood as a
level set of the induced probability field rather than as a decision boundary
learned by an independent classifier.

\begin{figure}[htbp]
\centering
\includegraphics[width=.9\textwidth]{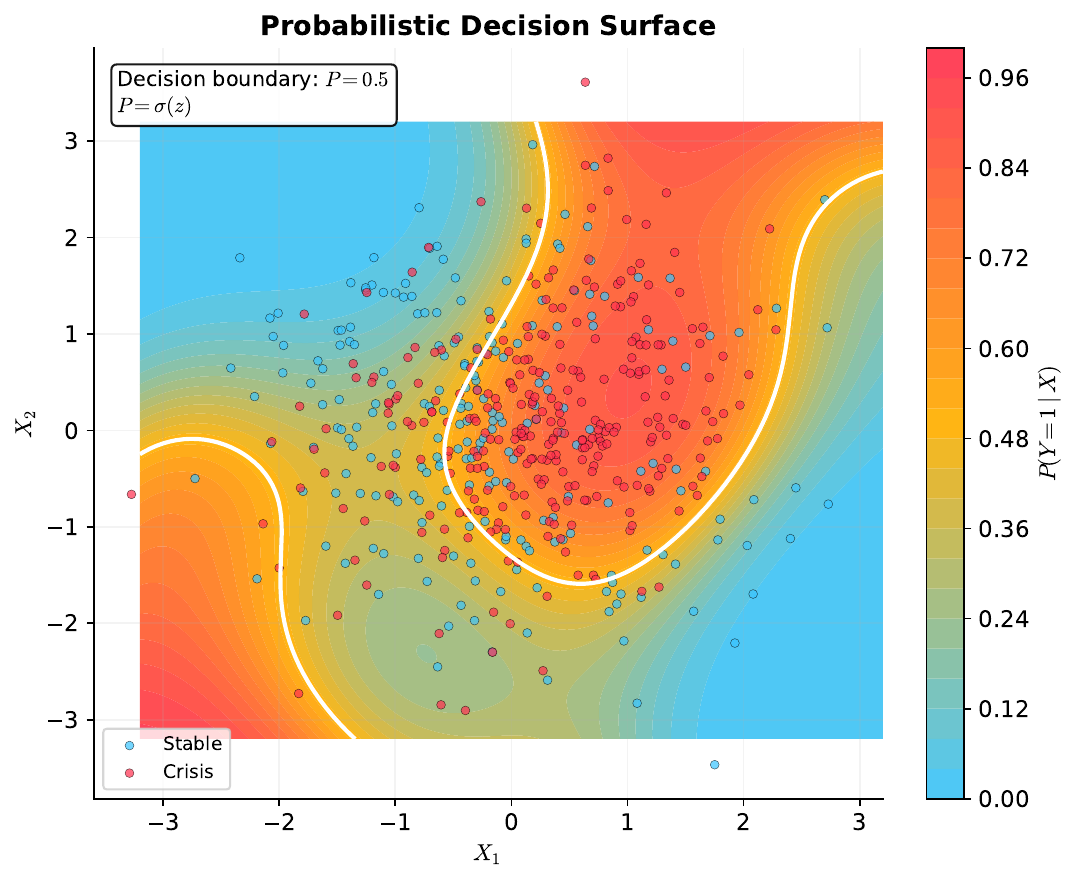}
\caption{
Probabilistic risk surface induced by the synthetic latent representation.
The colored background represents the crisis probability
$P(Y=1\mid X)=\sigma(f(X))$, while the white contour denotes the
probability level set $P(Y=1\mid X)=0.5$.
Stable and Crisis markers correspond to binary states sampled from the
associated Bernoulli distributions. The figure illustrates the continuous
transition from lower- to higher-risk regions generated by the probabilistic
scoring operator $\mathcal{P}$.
}
\label{fig:probability_surface}
\end{figure}

Figure~\ref{fig:probability_surface} shows how the nonlinear latent risk
representation is transformed into a continuous probability field over the
synthetic feature space. For each observation, the probabilistic scoring
operator maps the latent value $z_i$ into

\[
p_i=\sigma(z_i),
\]

so that larger latent values correspond monotonically to larger crisis
probabilities.

The contour $P(Y=1\mid X)=0.5$ should not be interpreted as a decision
boundary learned from data. Because the logistic function satisfies
$\sigma(0)=0.5$, this contour is equivalently the level set

\[
\left\{
X\in\mathcal{X} :
P(Y=1\mid X)=0.5
\right\}
=
\left\{
X\in\mathcal{X} :
f(X)=0
\right\}.
\]

It therefore provides a geometric reference separating regions with
probabilistic scores below and above $0.5$, while preserving the continuous
character of the underlying probability field. Observations near the contour
are associated with probabilities close to $0.5$, rather than with
classification uncertainty produced by a trained predictive model.

Together, Figures~\ref{fig:latent_space} and~\ref{fig:probability_surface}
illustrate the first two transformations of the operator pipeline,

\[
\mathcal{X}
\xrightarrow{\mathcal{D}}
\mathcal{L}
\xrightarrow{\mathcal{P}}
\mathcal{R},
\]

from synthetic observations to a nonlinear latent representation and then
to probabilistic risk scores. The subsequent stages introduce network
structure and time-dependent contagion, allowing these probabilistic states
to be examined within the dynamic visual analytics framework.

\begin{figure}[htbp]
\centering
\includegraphics[width=.99\textwidth]{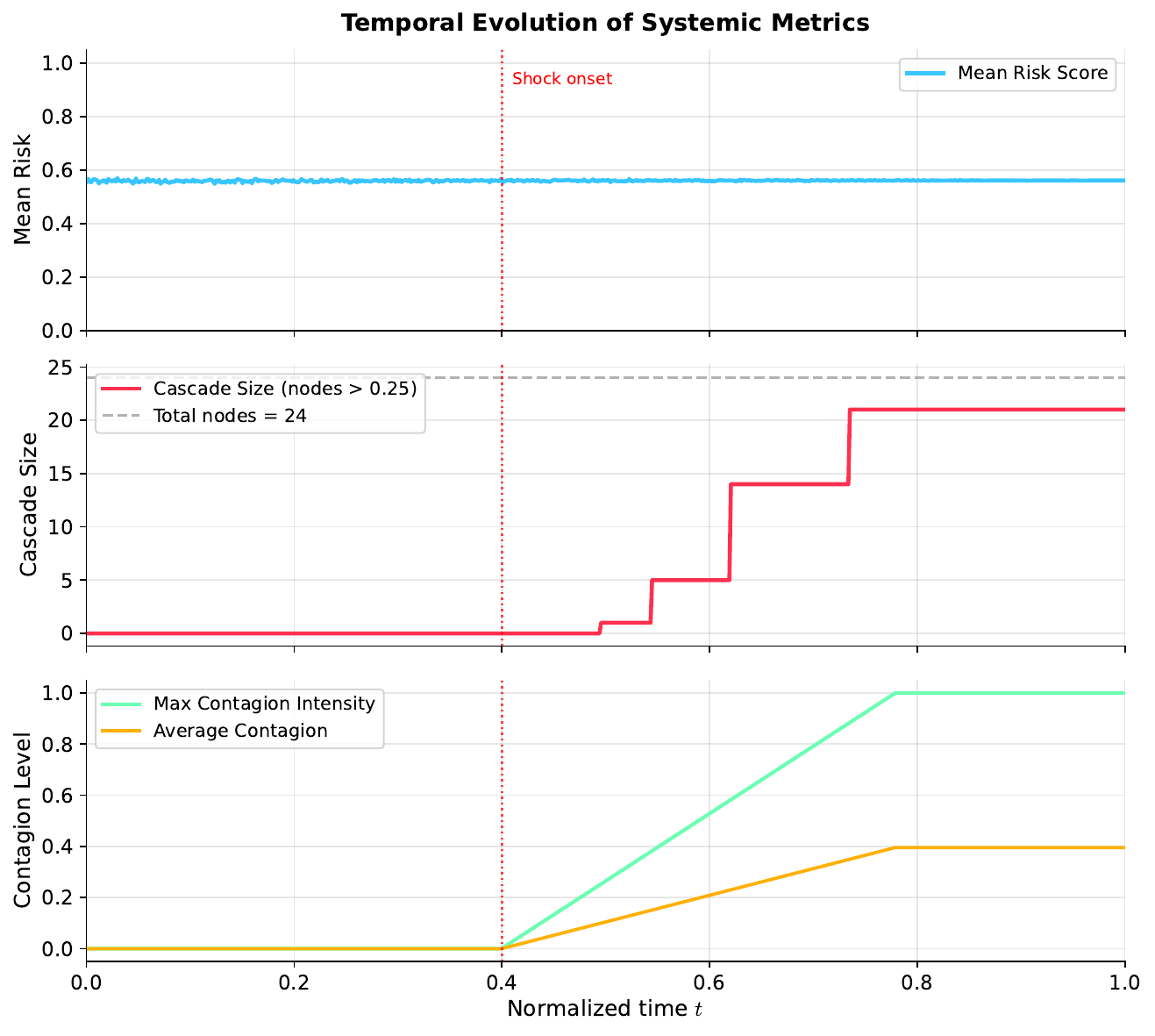}
\caption{
Temporal evolution of the principal quantitative indicators in the synthetic
systemic-risk experiment. The upper panel shows the mean probabilistic risk
$\bar{p}(t)$. The middle panel reports the cascade size $K(t)$, defined as
the number of nodes whose contagion intensity exceeds the prescribed
threshold $\tau$. The lower panel shows the maximum contagion intensity
$C_{\max}(t)$ and the network-average contagion intensity $\bar{C}(t)$.
Together, these quantities distinguish the evolution of the probabilistic
risk field from the propagation of network-mediated contagion.
}
\label{fig:systemic_metrics}
\end{figure}

Figure~\ref{fig:systemic_metrics} summarizes the temporal evolution of the
main quantitative indicators used to characterize the synthetic experiment.
The mean probabilistic risk is defined as

\begin{equation}
\bar{p}(t)
=
\frac{1}{N}
\sum_{i=1}^{N} p_i(t),
\label{eq:mean_probability}
\end{equation}

providing an aggregate measure of the time-dependent probabilistic scores
generated by the operator $\mathcal{P}$.

Network-mediated propagation is characterized separately through the
cascade size

\begin{equation}
K(t)
=
\sum_{i=1}^{M}
\mathbf{1}
\left\{
C_i(t)>\tau
\right\},
\label{eq:cascade_size}
\end{equation}

where $\tau$ denotes the contagion threshold. Thus, $K(t)$ counts the
number of network nodes whose contagion intensity exceeds the prescribed
threshold at time $t$.

Two additional indicators summarize the magnitude of the contagion process:

\begin{equation}
C_{\max}(t)
=
\max_{1\leq i\leq M} C_i(t),
\label{eq:max_contagion}
\end{equation}

and

\begin{equation}
\bar{C}(t)
=
\frac{1}{M}
\sum_{i=1}^{M} C_i(t).
\label{eq:mean_contagion}
\end{equation}

The temporal profiles in Figure~\ref{fig:systemic_metrics} reveal a
distinction between the probabilistic risk component and the network
contagion component. In the present synthetic configuration,
$\bar{p}(t)$ remains comparatively stable over time, whereas the onset of
the shock produces a progressive increase in cascade size and contagion
intensity. The increase in $K(t)$ occurs in discrete steps because nodes
cross the contagion threshold as the shock propagates through successive
shortest-path distances from the source node.

This behavior illustrates an important feature of the operator-based
architecture: temporal variation in network-mediated contagion can be
examined separately from variation in the underlying probabilistic risk
scores. The visualization therefore exposes the evolution of distinct
quantitative states without interpreting them as predictive-performance
metrics.

\begin{figure}[htbp]
\centering
\includegraphics[width=.82\textwidth]{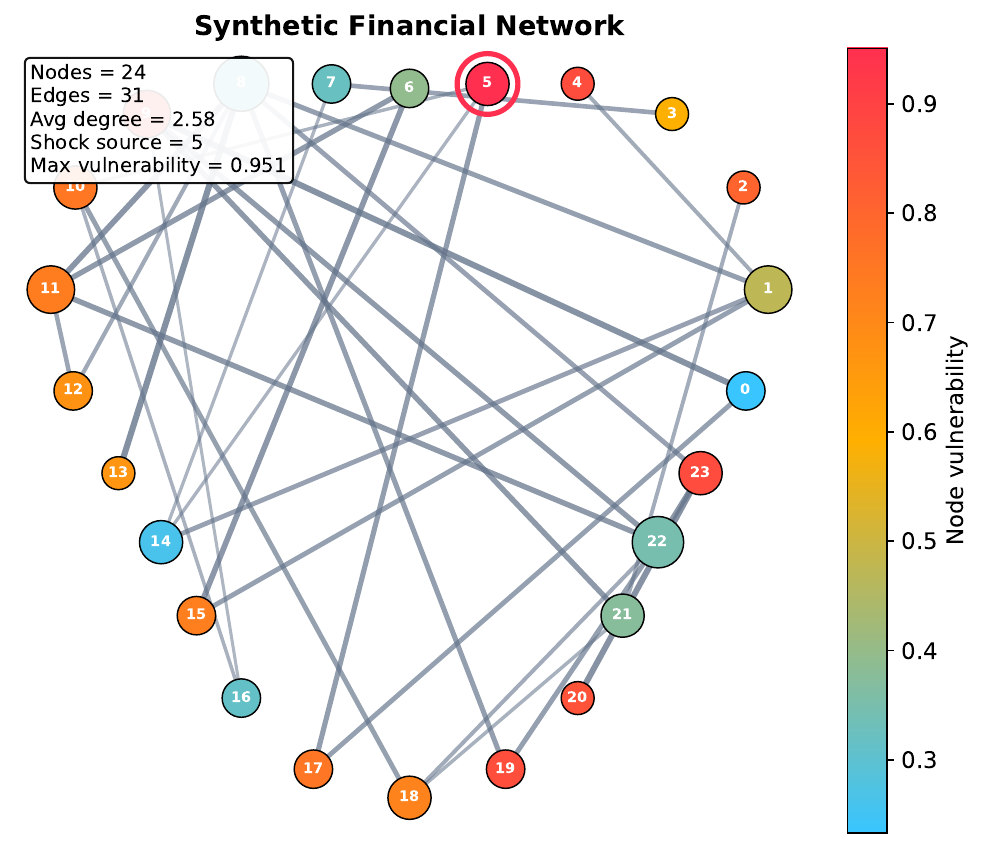}
\caption{
Synthetic weighted financial network used in the contagion experiment.
Node color encodes the assigned vulnerability score $\nu_i$, while node
size represents node degree. Edge thickness represents the corresponding
network weight $w_{ij}$. The highlighted node denotes the initial shock
source $v_0$, from which shortest-path distances are computed for the
subsequent contagion dynamics.
}
\label{fig:network}
\end{figure}

Figure~\ref{fig:network} illustrates the network structure generated by the
operator $\mathcal{G}$. Each node $v_i\in V$ is associated with an assigned
vulnerability score $\nu_i$, while the weighted edge set represents the
synthetic connectivity structure used in the experiment. These node-level
and topological quantities are maintained as distinct components of the
network representation.

The initial shock source $v_0$ provides the reference node for the contagion
process. For every node reachable from $v_0$, the shortest-path distance

\[
d_i=d_G(v_i,v_0)
\]

determines the topological separation from the source. The contagion operator
$\mathcal{C}$ then maps this distance into the time-dependent intensity

\[
C_i(t)=S(t)\exp(-\lambda d_i),
\]

as defined in Eq.~(\ref{eq:contagion_intensity}). Consequently, contagion
intensity decreases exponentially with graph distance for a fixed shock
intensity $S(t)$.

The network representation also distinguishes intrinsic vulnerability from
topological position. A node with a large vulnerability score $\nu_i$ need
not have a large degree, and a highly connected node need not have a large
vulnerability score. Likewise, vulnerability and contagion intensity remain
distinct quantities: $\nu_i$ is an assigned node attribute, whereas $C_i(t)$
is generated dynamically from the shock process and the node's shortest-path
distance from $v_0$.

This separation allows the framework to represent node-level heterogeneity
and network-mediated propagation without conflating the two mechanisms.
The network generated by $\mathcal{G}$ therefore provides the structural
domain on which the contagion operator $\mathcal{C}$ subsequently acts.

\begin{figure}[htbp]
\centering
\includegraphics[width=\textwidth]{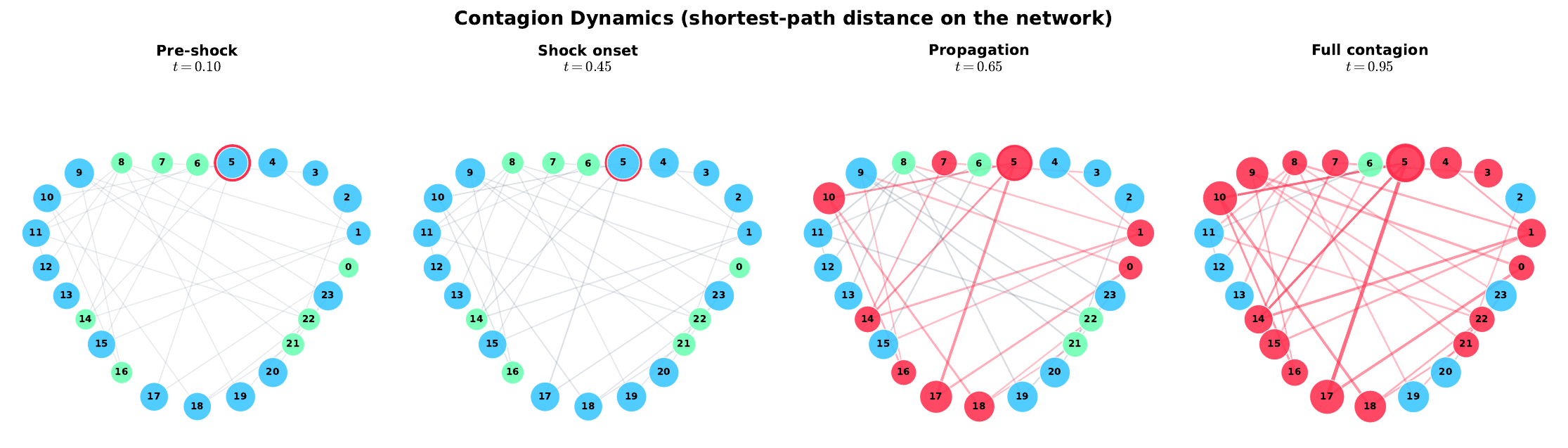}
\caption{
Temporal evolution of the synthetic contagion process at representative
simulation times. From left to right, the panels illustrate the pre-shock
state, shock onset, intermediate propagation, and the final contagion state.
Node color encodes the time-dependent contagion intensity $C_i(t)$, while
the propagation pattern reflects shortest-path distance from the initial
shock source $v_0$. The sequence illustrates how increasing shock intensity
produces progressively broader network activation under the prescribed
distance-attenuation mechanism.
}
\label{fig:contagion}
\end{figure}

Figure~\ref{fig:contagion} illustrates the temporal evolution of the
synthetic shock-propagation process. For each node $v_i$ reachable from the
initial shock source $v_0$, contagion intensity is determined by

\[
C_i(t)
=
S(t)\exp\!\left[-\lambda d_G(v_i,v_0)\right],
\]

as defined in Eq.~(\ref{eq:contagion_intensity}). As the normalized shock
intensity $S(t)$ increases, nodes at progressively larger shortest-path
distances from the source attain higher contagion intensities, producing
the expanding activation pattern observed across the panels.

For a fixed time $t$, the effect of network distance can be expressed as

\[
\frac{C_i(t)}{S(t)}
=
\exp(-\lambda d_i),
\]

showing explicitly that the contribution of the shock decays exponentially
with shortest-path distance $d_i=d_G(v_i,v_0)$. The parameter $\lambda$
therefore controls the rate of topological attenuation: larger values of
$\lambda$ produce faster decay with graph distance, whereas smaller values
allow the shock to retain greater intensity at more distant nodes.

The sequence also illustrates the distinction between temporal shock
amplification and topological attenuation. The function $S(t)$ governs the
overall magnitude of the disturbance through time, while
$\exp(-\lambda d_i)$ determines how that disturbance is distributed across
the network relative to the source node. Their product generates the
time-dependent contagion field visualized in
Figure~\ref{fig:contagion}.

\begin{figure}[htbp]
\centering
\includegraphics[width=\textwidth]{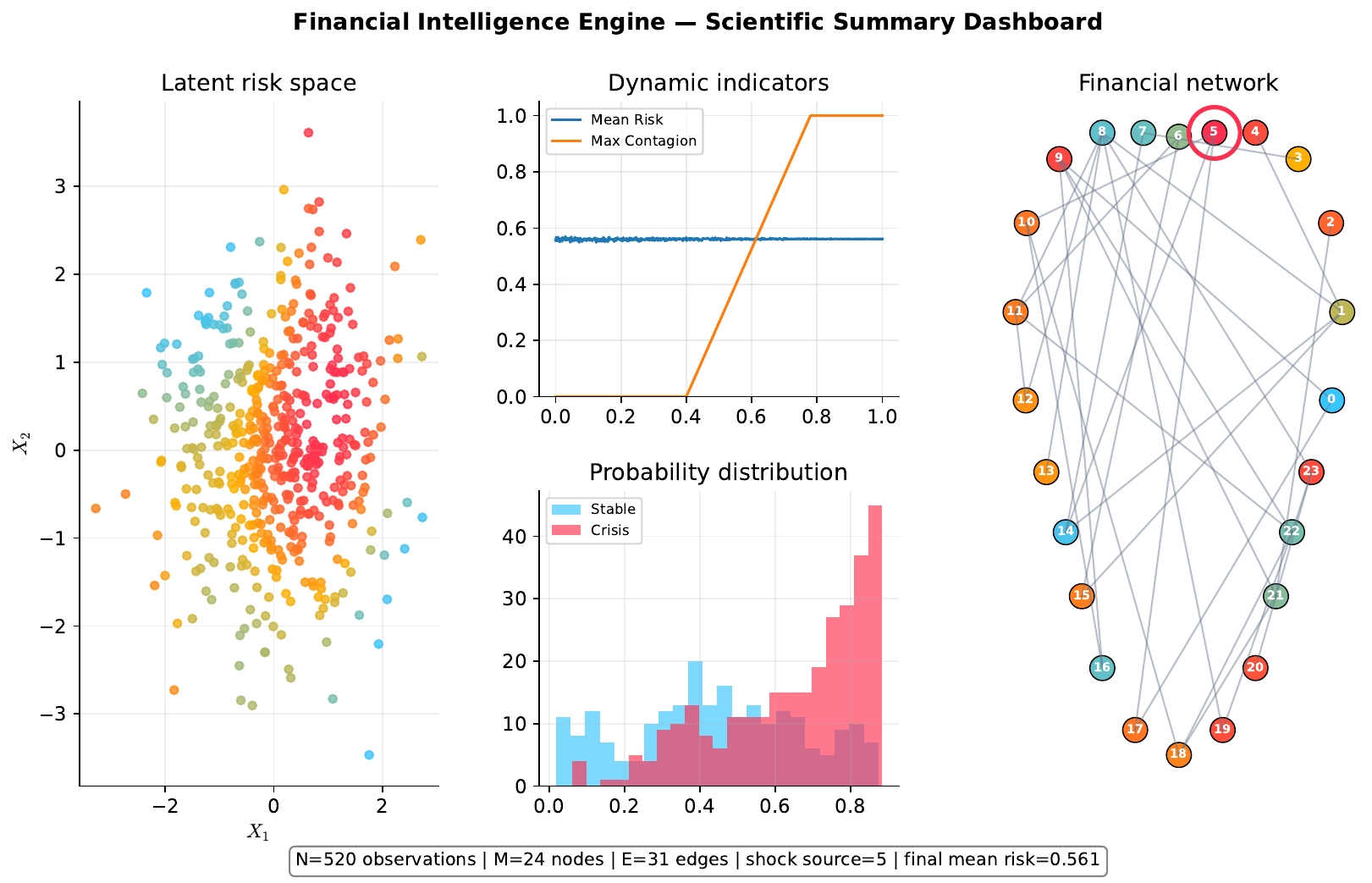}
\caption{
Integrated scientific dashboard summarizing the synthetic experiment.
The visualization combines the latent risk representation, temporal
systemic-risk indicators, probabilistic risk distribution, and weighted
financial-network structure within a common analytical view. The panels
represent complementary states of the same operator-based pipeline rather
than independent computational outputs.
}
\label{fig:dashboard}
\end{figure}

\begin{figure}[htbp]
\centering

\begin{tikzpicture}[
    scale=0.92,
    transform shape,
    node distance=2.2cm,
    every node/.style={font=\small},
    box/.style={
        draw=blue!70!black,
        rounded corners=5pt,
        minimum width=3.5cm,
        minimum height=1.08cm,
        align=center,
        fill=blue!4,
        thick
    },
    pipeline/.style={
        draw=red!70!black,
        rounded corners=6pt,
        minimum width=5.0cm,
        minimum height=1.35cm,
        align=center,
        fill=red!8,
        very thick
    },
    arrow/.style={
        ->,
        thick,
        >=Stealth
    }
]


\node[pipeline] (PIPE) at (0,0) {
    \textbf{Operator-Based Visual Analytics Pipeline}\\[1.2mm]
    $\mathcal{F}
    =
    \Pi
    \circ
    \Psi
    \circ
    \mathcal{C}
    \circ
    \mathcal{G}
    \circ
    \mathcal{P}
    \circ
    \mathcal{D}$
};


\node[box] (PROB) at (0,3.1) {
    \textbf{Probabilistic Modeling}\\
    Synthetic Risk Scoring
};

\node[box] (NET) at (-5.4,0) {
    \textbf{Network Science}\\
    Financial Networks
};

\node[box] (VIS) at (5.4,0) {
    \textbf{Visual Analytics}\\
    Scientific Visualization
};

\node[box] (COMP) at (0,-3.1) {
    \textbf{Scientific Computing}\\
    Numerical Simulation
};


\draw[arrow] (PROB.south) -- (PIPE.north);
\draw[arrow] (NET.east) -- (PIPE.west);
\draw[arrow] (VIS.west) -- (PIPE.east);
\draw[arrow] (COMP.north) -- (PIPE.south);

\end{tikzpicture}

\caption{%
Conceptual positioning of the proposed operator-based visual analytics
pipeline. The architecture combines probabilistic modeling, network science,
visual analytics, and scientific computing within a unified operator
formulation. The present synthetic implementation uses probabilistic risk
scoring rather than a trained predictive model, while the network and
contagion components provide the structural and dynamic states subsequently
mapped into scientific visual representations.
}
\label{fig:operator_architecture}

\end{figure}
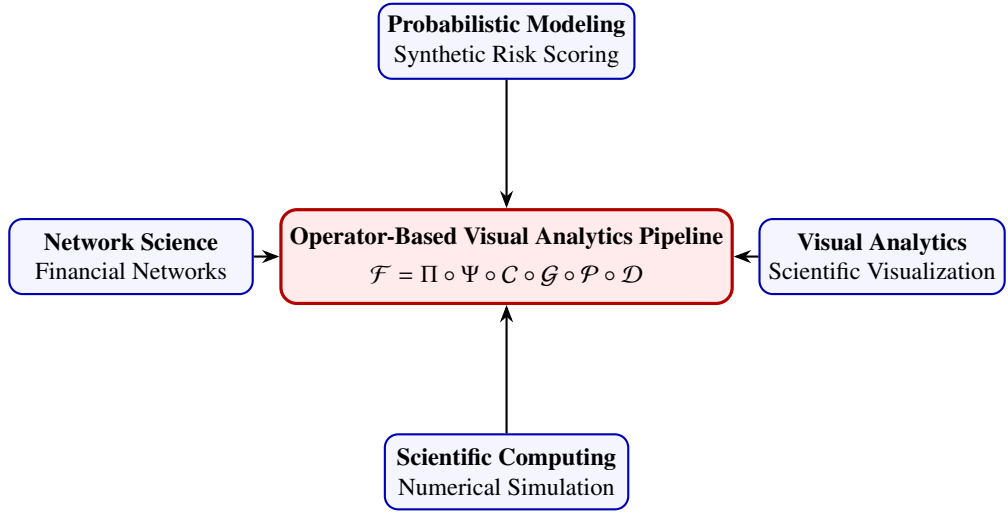

Figure~\ref{fig:operator_architecture} provides a complementary conceptual
view of the proposed architecture. Whereas
Figure~\ref{fig:fie_architecture} describes the sequential computational
workflow

\[
\mathcal{D}
\rightarrow
\mathcal{P}
\rightarrow
\mathcal{G}
\rightarrow
\mathcal{C}
\rightarrow
\Psi
\rightarrow
\Pi,
\]

Figure~\ref{fig:operator_architecture} emphasizes the methodological domains
that contribute to the pipeline.

Probabilistic modeling provides the synthetic risk-scoring component;
network science supplies the graph representation and shortest-path structure
used by the contagion mechanism; scientific computing supports the numerical
simulation of the time-dependent states; and visual analytics provides the
mapping and coordinated representation of the resulting quantitative
information.

The diagram therefore positions the proposed architecture as an
interdisciplinary computational pipeline without implying that the present implementation introduces a new machine-learning, network, or contagion model.

Figures~\ref{fig:latent_space}--\ref{fig:operator_architecture} collectively illustrate the computational realization of the proposed operator-based visual analytics pipeline. Rather than representing independent outputs, the figures provide complementary views of successive stages of the same computational architecture, progressing from latent risk representation and probabilistic scoring to network construction, contagion dynamics, and integrated scientific visualization.

The relevance of the dashboard lies not merely in the simultaneous display of multiple panels, but in the preservation of the relationships among the quantitative states generated throughout the pipeline. The latent-space and probability panels characterize the synthetic probabilistic component, the temporal indicators summarize the evolution of probabilistic risk and network-mediated contagion, and the network panel exposes the structural domain on which the contagion process operates.

From the operator perspective, the dashboard therefore provides an
integrated representation of quantities generated at different stages of

\[
\mathcal{D}
\rightarrow
\mathcal{P}
\rightarrow
\mathcal{G}
\rightarrow
\mathcal{C}
\rightarrow
\Psi
\rightarrow
\Pi.
\]

This coordinated representation facilitates comparison among probabilistic, temporal, and topological quantities while preserving their distinct mathematical roles within the framework.

\section{Experimental Configuration and Results}
\label{sec:experiments}

This section describes the experimental configuration and summarizes the quantitative behavior of the proposed operator-based visual analytics pipeline. All experiments are conducted using synthetic data and are intended to evaluate the computational behavior and internal consistency of the pipeline rather than its predictive performance on empirical financial data.

For reproducibility, the following tables report the mathematical operators, simulation parameters, and structural characteristics of the synthetic financial network used throughout the experiment.

\begin{table}[H]
\centering
\caption{Mathematical operators and their computational roles in the proposed operator-based visual analytics pipeline.}
\label{tab:operators}

\begin{tabular}{lll}
\toprule
Operator & Mathematical role & Implementation \\
\midrule
$\mathcal{D}$ &
Latent risk representation &
Nonlinear synthetic risk surface \\

$\mathcal{P}$ &
Probabilistic risk scoring &
Logistic transformation \\

$\mathcal{G}$ &
Financial-network construction &
Weighted synthetic graph \\

$\mathcal{C}$ &
Dynamic contagion process &
Shortest-path-based propagation \\

$\Psi$ &
Visual mapping &
Position, size, opacity, and color encoding \\

$\Pi$ &
Geometric projection &
Two-dimensional visual representation \\
\bottomrule
\end{tabular}

\end{table}

Table~\ref{tab:operators} summarizes the correspondence between the
mathematical operators introduced in Section~2 and their computational realization in the synthetic experiment. Each operator performs a distinct transformation within the sequential information flow

\[
\mathcal{D}
\rightarrow
\mathcal{P}
\rightarrow
\mathcal{G}
\rightarrow
\mathcal{C}
\rightarrow
\Psi
\rightarrow
\Pi.
\]

The operators are therefore not interpreted as independent analytical
outputs, but as compatible stages of a common computational architecture. Synthetic observations are transformed into a latent risk representation and probabilistic scores, associated with a weighted network, propagated through the prescribed contagion mechanism, and finally mapped into visual attributes and projected onto the visualization domain.

This modular organization also separates the architecture of the pipeline from the specific mechanisms adopted in the present proof-of-concept.
Alternative probabilistic scoring rules, network-construction procedures, contagion mechanisms, or visual encodings may replace the corresponding operators provided that compatibility between consecutive input and output representations is preserved.

\begin{table}[H]
\centering
\caption{Reference configuration of the synthetic numerical experiment.}
\label{tab:experimental_configuration}
\begin{tabular}{lll}
\hline
\textbf{Parameter} & \textbf{Description} & \textbf{Value} \\
\hline
$N$ & Number of synthetic observations & 520 \\
$M$ & Number of network nodes & 24 \\
$E$ & Number of network edges & 31 \\
$T$ & Number of simulation time steps & 672 \\
$\lambda$ & Contagion distance-decay parameter & 0.42 \\
$\tau$ & Cascade activation threshold & 0.25 \\
$t_0$ & Normalized shock-onset time & 0.40 \\
$\Delta t$ & Shock-growth interval & 0.38 \\
$v_0$ & Initial shock source & 5 \\
$\bar p(T)$ & Final mean probabilistic risk & 0.561 \\
\hline
\end{tabular}
\end{table} 
Table~\ref{tab:experimental_configuration} reports the reference configuration used in the synthetic experiment. The probabilistic component contains $N=520$ synthetic observations, while the network component consists of $M=24$ nodes and $E=31$ weighted edges. Node $5$ is used as the initial shock source for the contagion experiment.

The parameter values define a single controlled reference realization
and were selected to make the successive stages of the operator pipeline observable within a common simulation horizon. They were not obtained through calibration, optimization, or fitting to empirical financial data.
Accordingly, the reported numerical values should be interpreted as
properties of the reference synthetic experiment rather than as estimated characteristics of a real financial system.

The final mean probabilistic risk,

\begin{equation}
\bar{p}(T)
=
\frac{1}{N}
\sum_{i=1}^{N} p_i(T),
\end{equation}

is approximately $0.561$ for the reported realization. This quantity
summarizes the probabilistic risk scores at the final simulation time and should not be interpreted as a predictive-performance measure.

The configuration in Table~\ref{tab:experimental_configuration} represents a reference realization of the operator-based pipeline rather than an optimized parameterization. The experiment is designed to illustrate the interaction among probabilistic risk scoring, network structure, contagion dynamics, and visual representation under controlled synthetic conditions.

The node-level configuration provides an additional distinction between assigned vulnerability and network connectivity. These quantities represent different properties of the synthetic network and are therefore reported separately in Table~\ref{tab:network_nodes}.

\begin{table}[H]
\centering
\caption{Node-level structural properties of the synthetic financial network.}
\label{tab:network_nodes}

\begin{tabular}{rrrr}
\toprule
Node & Sector & Vulnerability & Degree \\
\midrule
0  & 1 & 0.232873 & 2 \\
1  & 0 & 0.471233 & 4 \\
2  & 2 & 0.803000 & 1 \\
3  & 0 & 0.590042 & 1 \\
4  & 0 & 0.869218 & 1 \\
5  & 0 & 0.950602 & 3 \\
6  & 0 & 0.392681 & 2 \\
7  & 0 & 0.322116 & 2 \\
8  & 2 & 0.304862 & 6 \\
9  & 1 & 0.884456 & 4 \\
10 & 0 & 0.750371 & 3 \\
11 & 1 & 0.733139 & 4 \\
12 & 2 & 0.676433 & 2 \\
13 & 2 & 0.665270 & 1 \\
14 & 0 & 0.262257 & 3 \\
15 & 0 & 0.727356 & 2 \\
16 & 0 & 0.313709 & 2 \\
17 & 1 & 0.754225 & 2 \\
18 & 2 & 0.715620 & 3 \\
19 & 1 & 0.863204 & 2 \\
20 & 0 & 0.851850 & 1 \\
21 & 2 & 0.375748 & 3 \\
22 & 2 & 0.346118 & 5 \\
23 & 2 & 0.871636 & 3 \\
\bottomrule
\end{tabular}

\end{table}

Table~\ref{tab:network_nodes} reports the sector assignment, vulnerability score $\nu_i$, and degree $k_i$ of each node in the reference network realization. The network contains $M=24$ nodes and $E=31$ weighted edges, corresponding to an average degree

\begin{equation}
\bar{k}
=
\frac{1}{M}\sum_{i=1}^{M}k_i
=
\frac{2E}{M}
=
\frac{62}{24}
\approx 2.58.
\label{eq:average_degree}
\end{equation}

Vulnerability and degree are treated as distinct node attributes.
The vulnerability score $\nu_i$ is assigned by the synthetic node-generation mechanism, whereas the degree $k_i$ is determined by the realized network topology. Consequently, a node with high vulnerability need not be among the most highly connected nodes, and conversely a node with high degree need not exhibit high vulnerability.

In the reference realization, node~5 is used as the initial shock source $v_0$. Its selection determines the origin from which shortest-path distances $d_G(v_i,v_0)$ are computed and therefore provides the initial condition for the contagion operator $\mathcal{C}$. The reported maximum vulnerability in the network is approximately $0.951$.

Taken together, Tables~\ref{tab:operators}--\ref{tab:network_nodes}
document the principal components of the reference synthetic experiment.
Table~\ref{tab:operators} identifies the mathematical operators and their computational roles, Table~\ref{tab:experimental_configuration} specifies the numerical
configuration, and Table~\ref{tab:network_nodes} reports the node-level properties of the realized network. These quantities provide the numerical context required to interpret the probabilistic, topological, and contagion results presented in the accompanying figures.

\section{Discussion}
\label{sec:discussion}

\subsection{Operator-Based Integration}

The results illustrate the role of operator composition as an organizational principle for the synthetic systemic-risk experiment. Rather than treating risk scoring, network construction, contagion dynamics, and visualization as independent computational procedures, the proposed architecture represents them as successive transformations,

\[
\mathcal{F}
=
\Pi
\circ
\Psi
\circ
\mathcal{C}
\circ
\mathcal{G}
\circ
\mathcal{P}
\circ
\mathcal{D}.
\]

This formulation makes explicit both the ordering of the computational stages and the information transferred between them. The latent representation operator $\mathcal{D}$ generates a nonlinear synthetic risk signal;
$\mathcal{P}$ maps this signal into probabilistic risk scores;
$\mathcal{G}$ introduces network structure; $\mathcal{C}$ generates
time-dependent contagion states; and $\Psi$ and $\Pi$ transform these
quantitative states into visual representations.

The principal advantage of this formulation is therefore architectural rather than predictive. The present implementation does not introduce a trained machine-learning model, nor does it propose a new financial contagion mechanism. Instead, it demonstrates how heterogeneous computational components
can be represented within a common mathematical pipeline while retaining their individual interpretations.

\subsection{Probabilistic Risk and Network-Mediated Contagion}

An important feature of the experiment is the separation between probabilistic risk scores and network-mediated contagion. The probabilistic component produces time-dependent scores $p_i(t)$ from the synthetic latent representation, whereas the contagion component operates on the graph through

\[
C_i(t)
=
S(t)\exp\!\left[-\lambda d_G(v_i,v_0)\right].
\]

These quantities describe different aspects of the synthetic system.
The probabilistic score characterizes the state associated with an observation, while contagion intensity depends explicitly on the temporal shock magnitude and the topological distance from the shock source.

This distinction is visible in the temporal indicators reported in
Figure~\ref{fig:systemic_metrics}. In the reference experiment, the mean probabilistic risk remains comparatively stable, while cascade size and contagion intensity respond more strongly to the evolution of the shock.
The observed behavior illustrates that changes in network-mediated propagation need not be interpreted as equivalent changes in the underlying probabilistic risk field.

The network experiment further separates assigned vulnerability from
topological connectivity. Vulnerability $\nu_i$, node degree $k_i$, and contagion intensity $C_i(t)$ represent distinct quantities. This distinction is relevant when interpreting synthetic propagation patterns because a node's assigned vulnerability does not by itself determine its structural position or its distance from the shock source.

\subsection{Relationship with Financial Network Models}

Financial-network models provide established mathematical mechanisms for studying systemic risk and contagion. Clearing models such as
Eisenberg--Noe \cite{Eisenberg2001}, threshold-based contagion models such as Gai--Kapadia \cite{gai2010}, and liquidity and asset-price feedback mechanisms \cite{lee2013,cifuentes2005} address substantially richer financial mechanisms than the distance-attenuation rule used in the present experiment.

The contagion operator introduced here should therefore be interpreted as a controlled synthetic mechanism for demonstrating the computational architecture, rather than as an alternative to these established models.
Its simple form makes the roles of shock intensity and graph distance
explicit, which is useful for examining how a contagion state propagates through the remaining stages of the operator pipeline.

A consequence of the modular formulation is that $\mathcal{C}$ is not tied to the particular exponential attenuation mechanism used in the present implementation. More sophisticated contagion mechanisms could replace this operator while preserving the surrounding architecture, provided that their output can be represented in a form compatible with the subsequent visual mapping operator $\Psi$.

\subsection{Visualization as an Explicit Computational Stage}

The visual component of the framework is represented explicitly through the operators $\Psi$ and $\Pi$. This separation distinguishes the transformation of quantitative states into graphical attributes from their final geometric projection.

The contribution should not be interpreted as claiming that visualization itself is novel as a mathematical or computational concept. Visual Analytics and scientific visualization already provide extensive methodologies for mapping complex quantitative information into interpretable visual representations \cite{thomas2005,keim2008}. The contribution of the present
framework is narrower: visual mapping is included explicitly within the same operator chain used to represent the synthetic risk, network, and contagion components.

This representation makes the dependency of the visualization on upstream computational states explicit. Changes in probabilistic scores, network structure, or contagion intensity propagate through $\Psi$ before the final projection $\Pi$ is produced. The resulting dashboard therefore represents coordinated views of quantities generated at different stages of the same computational pipeline rather than a collection of unrelated graphical outputs.

\subsection{Modularity and Reproducibility}

The operator formulation also provides a modular computational structure.
Each stage has a defined mathematical role and may, in principle, be replaced without requiring the complete architecture to be reformulated. For example, the logistic scoring operator could be replaced by an empirically estimated probabilistic model, the synthetic graph generator by an observed exposure
network, or the distance-based contagion operator by a clearing or
threshold-based mechanism.

Such replacements would change the substantive interpretation of the
experiment and would require independent validation. Modularity alone does not guarantee predictive validity or empirical realism. Its role is instead to make the computational dependencies explicit and to facilitate controlled comparison among alternative implementations.

Reproducibility similarly depends not only on the operator notation but on the complete specification of the experiment, including random seeds, network-generation rules, parameter values, simulation procedures, and software implementation. The numerical configuration and node-level properties reported in Section~\ref{sec:experiments} provide part of this specification, while the computational implementation provides the remaining procedural details.
This separation also permits reproducibility to be assessed at two
levels: numerical reproducibility of the intermediate computational
states and representational reproducibility of the visual outputs
generated from those states.

\subsection{Scope of the Contribution}

The contribution of the present study is best understood as a proof-of-concept operator-based visual analytics architecture for synthetic systemic-risk dynamics. It does not introduce a new predictive algorithm, establish empirical forecasting performance, or propose a new theory of financial contagion.

Within this scope, the experiments demonstrate three aspects of the
architecture: the transformation of a nonlinear latent signal into continuous probabilistic risk scores; the propagation of a controlled shock through a weighted network according to shortest-path distance; and the coordinated visual representation of probabilistic, temporal, and topological quantities.

The resulting framework provides a computational structure in which these components can be examined jointly while remaining mathematically distinct.
Its value therefore lies primarily in explicit composition, modularity, and the coordinated representation of heterogeneous systemic-risk quantities.

\section*{Code Availability}

The computational implementation, experimental configuration, random seeds, figure-generation routines, and source code used to generate the dynamic visualizations reported in this study are publicly available at:

\begin{center}
\url{https://github.com/IsabelCasPe/Papers-Publicados/blob/main/Financial_Intelligence_Engine_%28FIE%29.ipynb}
\end{center}

\section*{Supplementary Visualizations}

Two dynamic visualizations are provided as supplementary material to
illustrate the temporal behavior of the operator-based pipeline:

\begin{itemize}

    \item \textbf{Supplementary Visualization V1}: dynamic representation of
    the computational architecture, illustrating the coordinated evolution
    of probabilistic risk representations, network structure, and synthetic
    contagion states. Available at:
    \url{https://youtu.be/2KyQpRIPkko}

    \item \textbf{Supplementary Visualization V2}: dynamic implementation associated with the operator-based architecture and synthetic experiment analyzed in this manuscript. Available at:
    \url{https://www.youtube.com/watch?v=DCFy6AEs3KE}

\end{itemize}

The supplementary animations complement the static figures by displaying
the temporal evolution of quantities represented at different stages of the
pipeline. They are intended as visual complements to the quantitative
analysis presented in the manuscript and do not constitute additional
empirical validation.

\section{Conclusion}
\label{sec:conclusion}

This paper introduced an operator-based visual analytics pipeline for
exploring synthetic systemic-risk dynamics in financial networks. The
framework organizes latent risk representation, probabilistic risk scoring,
network construction, contagion dynamics, visual mapping, and geometric
projection within the unified operator composition

\[
\mathcal{F}
=
\Pi
\circ
\Psi
\circ
\mathcal{C}
\circ
\mathcal{G}
\circ
\mathcal{P}
\circ
\mathcal{D}.
\]

The principal contribution is architectural rather than predictive. The
present implementation does not introduce a trained machine-learning model
or a new financial contagion mechanism. Instead, it provides an explicit
mathematical representation of the information flow connecting synthetic
risk states, network-mediated propagation, and scientific visualization.

The numerical experiment demonstrated the computational realization of this
architecture under controlled synthetic conditions. Synthetic observations
were transformed into a nonlinear latent risk representation and subsequently
mapped into continuous probabilistic scores. A weighted financial network
provided the structural domain for a shortest-path-based contagion mechanism,
while temporal indicators characterized the evolution of cascade size and
contagion intensity. These quantitative states were subsequently transformed
into coordinated dynamic visual representations.

The experiment also highlighted the distinction among quantities that may
otherwise be conflated in systemic-risk visualization. Probabilistic risk,
assigned node vulnerability, network connectivity, and dynamic contagion
intensity were represented as distinct components of the pipeline. In
particular, the temporal analysis showed how network-mediated contagion can
evolve differently from the underlying aggregate probabilistic risk state
under the prescribed synthetic configuration.

A central property of the proposed formulation is modularity. Because the
operators are organized through compatible input and output representations,
individual components may be replaced by alternative implementations without
requiring the complete pipeline to be reformulated. The synthetic
probabilistic scoring mechanism may therefore be replaced by an empirically
estimated model, the reference network by observed financial exposures, and
the distance-based contagion rule by more sophisticated clearing, threshold,
liquidity, or multilayer mechanisms.

The framework also makes visual transformation explicit through the operators
$\Psi$ and $\Pi$. This formulation does not imply that visual mapping or
scientific visualization are themselves novel concepts. Rather, it makes
their relationship with upstream quantitative states explicit within the
same mathematical architecture used to describe the remaining computational
stages.

The scope of the present contribution is consequently that of a
proof-of-concept architecture demonstrated through a reproducible synthetic
experiment. Empirical validation, systematic sensitivity analysis, ablation
experiments, richer financial-network structures, and trained probabilistic
models remain necessary before the framework can support stronger claims
regarding real-world systemic-risk analysis.

Within this scope, the operator-based formulation provides a transparent
computational structure for composing heterogeneous analytical components
while preserving their distinct mathematical roles. Future work may extend
this architecture toward empirical financial systems, alternative contagion
mechanisms, data-driven probabilistic models, temporal and multilayer
networks, and interactive visual analytics.


\begin{thebibliography}{99}

\bibitem[Acemoglu et al.(2015)]{acemoglu2015}
\newblock Acemoglu D, Ozdaglar A, Tahbaz-Salehi A (2015)
\newblock Systemic Risk and Stability in Financial Networks.
\newblock \emph{American Economic Review} 105(2):564--608.

\bibitem[Barabási(2016)]{barabasi2016}
\newblock Barabási AL (2016)
\newblock Network Science.
\newblock Cambridge University Press.

\bibitem[Battiston et al.(2012)]{battiston2012}
\newblock Battiston S, Puliga M, Kaushik R, Tasca P, Caldarelli G (2012)
\newblock DebtRank: Too Central to Fail? Financial Networks, the FED and Systemic Risk.
\newblock \emph{Scientific Reports} 2:541.

\bibitem[Battiston et al.(2016)]{Battiston2016}
\newblock Battiston S, Caldarelli G, May RM, Roukny T, Stiglitz JE (2016)
\newblock The Price of Complexity in Financial Networks.
\newblock \emph{Proceedings of the National Academy of Sciences} 113(36):10031--10036.

\bibitem[Bishop(2006)]{bishop2006}
\newblock Bishop CM (2006)
\newblock Pattern Recognition and Machine Learning.
\newblock Springer.

\bibitem[Černevičienė and Kabašinskas(2024)]{cerneviciene2024}
\newblock Černevičienė J, Kabašinskas A (2024)
\newblock Explainable Artificial Intelligence (XAI) in Finance:
A Systematic Literature Review.
\newblock \emph{Artificial Intelligence Review} 57:216.
\newblock doi:10.1007/s10462-024-10854-8.

\bibitem[Cifuentes et al.(2005)]{cifuentes2005}
\newblock Cifuentes R, Ferrucci G, Shin HS (2005)
\newblock Liquidity Risk and Contagion.
\newblock \emph{Journal of the European Economic Association} 3(2--3):556--566.

\bibitem[Eisenberg and Noe(2001)]{Eisenberg2001}
\newblock Eisenberg L, Noe TH (2001)
\newblock Systemic Risk in Financial Systems.
\newblock \emph{Management Science} 47(2):236--249.

\bibitem[Gai and Kapadia(2010)]{gai2010}
\newblock Gai P, Kapadia S (2010)
\newblock Contagion in Financial Networks.
\newblock \emph{Proceedings of the Royal Society A} 466(2120):2401--2423.

\bibitem[Gonon et al.(2026)]{gonon2026}
\newblock Gonon L, Meyer-Brandis T, Weber N (2026)
\newblock Computing Systemic Risk Measures with Graph Neural Networks.
\newblock \emph{SIAM Journal on Financial Mathematics} 17(2):565--619.
\newblock doi:10.1137/24M1697402.

\bibitem[Goodfellow et al.(2016)]{goodfellow2016}
\newblock Goodfellow I, Bengio Y, Courville A (2016)
\newblock Deep Learning.
\newblock MIT Press.

\bibitem[Guidotti et al.(2018)]{guidotti2018}
\newblock Guidotti R, Monreale A, Ruggieri S, Turini F, Giannotti F, Pedreschi D (2018)
\newblock A Survey of Methods for Explaining Black Box Models.
\newblock \emph{ACM Computing Surveys} 51(5):93.

\bibitem[Hastie et al.(2009)]{hastie2009}
\newblock Hastie T, Tibshirani R, Friedman J (2009)
\newblock The Elements of Statistical Learning.
\newblock 2nd Edition.
\newblock Springer.

\bibitem[Keim et al.(2010)]{keim2008}
\newblock Keim DA, Kohlhammer J, Ellis G, Mansmann F (2010)
\newblock Mastering the Information Age: Solving Problems with Visual Analytics.
\newblock Eurographics Association.

\bibitem[Lee(2013)]{lee2013}
\newblock Lee SH (2013)
\newblock Systemic Liquidity Shortages and Interbank Network Dynamics.
\newblock \emph{Journal of Financial Stability} 9(1):1--12.

\bibitem[Munzner(2014)]{munzner2014}
\newblock Munzner T (2014)
\newblock Visualization Analysis and Design.
\newblock CRC Press.

\bibitem[Molnar(2022)]{molnar2022}
\newblock Molnar C (2022)
\newblock Interpretable Machine Learning.
\newblock 2nd Edition.
\newblock Lulu Press.

\bibitem[Newman(2010)]{newman2010}
\newblock Newman MEJ (2010)
\newblock Networks: An Introduction.
\newblock Oxford University Press. 

\bibitem[Pacelli et al.(2025)]{pacelli2025}
\newblock Pacelli V, Panetta IC, Povia MM (2025)
\newblock Systemic Risk and Network Science: A Bibliometric and Systematic Review.
\newblock In: Pacelli V (ed), \emph{Systemic Risk and Complex Networks in Modern Financial Systems}.
\newblock New Economic Windows, Springer, Cham, pp. 21--42.
\newblock doi:10.1007/978-3-031-64916-5\_2.

\bibitem[Pang and Fan(2024)]{pangfan2024}
\newblock Pang C, Fan H (2024)
\newblock Risk Amplification Effect of Multilayer Financial Networks:
Feedback Mechanism or Cyclic Structure?
\newblock \emph{Economics Letters} 242:111887.
\newblock doi:10.1016/j.econlet.2024.111887.

\bibitem[Pereda(2025)]{pereda2025}
\newblock Pereda AIC (2025)
\newblock Systemic Risk and Default Cascades in Global Equity Markets:
Extending the Gai--Kapadia Framework with Stochastic Simulations and Network Analysis.
\newblock arXiv:2504.01969.

\bibitem[Pereda(2026)]{pereda2026}
\newblock Pereda AIC (2026)
\newblock Systemic Risk and Default Cascades in Global Equity Markets:
A Network and Tail-Risk Approach Based on the Gai--Kapadia Framework.
\newblock arXiv:2604.19796.

\bibitem[Castillo Pereda(2026)]{castillopereda2026jaes}
\newblock Castillo Pereda AI (2026)
\newblock Systemic Risk and Contagion in Global Equity Markets:
An Applied Economic Analysis of Emerging Market Cascades.
\newblock \emph{Journal of Applied Economic Sciences}
21(3(93)):809--835.
\newblock doi:10.57017/jaes.v21.3(93).07.

\bibitem[Purnell et al.(2024)]{purnell2024}
\newblock Purnell D Jr, Etemadi A, Kamp J (2024)
\newblock Developing an Early Warning System for Financial Networks:
An Explainable Machine Learning Approach.
\newblock \emph{Entropy} 26(9):796.
\newblock doi:10.3390/e26090796.

\bibitem[Press et al.(2007)]{numericalrecipes}
\newblock Press WH, Teukolsky SA, Vetterling WT, Flannery BP (2007)
\newblock Numerical Recipes: The Art of Scientific Computing.
\newblock 3rd Edition.
\newblock Cambridge University Press.

\bibitem[Thomas and Cook(2005)]{thomas2005}
\newblock Thomas JJ, Cook KA (2005)
\newblock Illuminating the Path: The Research and Development Agenda for Visual Analytics.
\newblock IEEE Computer Society.

\bibitem[Ware(2021)]{ware2021}
\newblock Ware C (2021)
\newblock Information Visualization: Perception for Design.
\newblock 4th Edition.
\newblock Morgan Kaufmann.

\bibitem[Yeo et al.(2025)]{yeo2025}
\newblock Yeo WJ, Van Der Heever W, Mao R, Cambria E, Satapathy R, Mengaldo G (2025)
\newblock A Comprehensive Review on Financial Explainable AI.
\newblock \emph{Artificial Intelligence Review} 58:189.
\newblock doi:10.1007/s10462-024-11077-7.

\end{thebibliography}
\end{document}